# Mathematical Modelling of Energy Wastage in Absence of Levelling & Sectoring in Wireless Sensor Networks

Priyanka Sharma, Garimella Rama Murthy

International Institute of Information Technology Hyderabad (IIIT H), India
(priyanka.sharma@research.iiit.ac.in; rammurthy@iiit.ac.in)

**Abstract.** In this paper, we quantitatively (mathematically) reason the energy savings achieved by the Leveling and Sectoring protocol. Due to the energy constraints on the sensor nodes (in terms of supply of energy) energy awareness has become crucial in networking protocol stack. The understanding of routing protocols along with energy awareness in a network would help in energy optimization with efficient routing .We provide analytical modelling of the energy wastage in the absence of Leveling and Sectoring protocol by considering the network in the form of binary tree, nested tree and Q-ary tree. The simulation results reflect the energy wastage in the absence of Levelling and Sectoring based hybrid protocol.

## 1      Introduction

Wireless Sensor Networks (WSNs) have been applied to a large number of disparate applications such as monitoring of environment, medical care and military sensing. As the networks find implementation in such a huge diversity of applications, the demand for efficiency has increased. The efficiency of sensor networks is defined in terms of network performance [2]. Regardless, the question of desirable performance depends upon a number of characteristics some of which are high throughput, network longevity and quality of service [10]. One of the important characteristic which has stimulated the interest is the energy efficient operation of the network. Incidentally, a major constraint in the case of Wireless Sensor Networks (WSNs) is the limited power supply. Hence, it becomes crucial that energy is used in an efficient manner without neglecting the output of the network. The monitoring of energy consumption will contribute in an optimal performance of the network.

The sensor nodes in a WSN are responsible for sensing the environmental information and reporting the same to the Base Station (BST). Hence, reliable routing of data is one of the major tasks in WSNs. Further, the energy constraints associated with the sensor nodes makes it important to ensure energy awareness at different layers of networking protocol stack [1].

Levelling and Sectoring is a routing protocol, proposed with the aim of improving network performance in WSNs [5]. As a part of comprehensive study, in this paper we estimate the wastage of energy when Levelling & Sectoring Protocol has not been implemented.

The rest of the paper has been divided as follows. Section 2 discusses the Background Work followed by the Motivation of the paper & Proposed Work in Section 3 and Section 4 respectively. Section 5 presents the Simulation Results. The paper is concluded in Section 6.

## 2      Background Work

Routing is difficult in WSNs due to the following reasons:

1. A sink is involved in the transmission of the sensed data from different sources.
2. Multiple sensors, present in the vicinity of the phenomena, leads to redundancy as they generate same data [1].

In Wireless Sensor Nodes, the power is stored in batteries which imply that the energy is available in a limited quantity. Hence, besides routing, the limited energy supply to the sensor nodes contribute to the challenges associated with the WSNs. In alignment with the challenge of reliable routing and energy efficiency, Levelling & Sectoring protocol aims towards the localization of the sensor node.

Routing in WSNs is performed through different techniques such as flooding, gossiping and directed controlled flooding [3]. These techniques may involve some transmissions which are not necessary. Levelling can be understood as the flow of information, directed towards the destination node. Assuming that the Base Station (BST) transmits signals at different power levels, all the nodes which receive the minimum power level signal from the BST set their level as 1 (This is the Level ID).Further, the signal power increases and all the nodes are assigned certain levels. Due to levelling, the strength of the signal related to the nodes is known. Based on this, sectors are created with equal angles. The destination node is in the center and the sectors are made either clockwise or anticlockwise. Once the sectors have been made, BST along with its local information sends Sector Ids. Once the Level Ids and Sector Ids have been assigned every node in the network has a {node_id: level_id; sensor_id}. The BST broadcasts a content based query containing {data_type; data_operator; data_threshold}. Route Reply {node_id; level_id; sector_id; data_type; data_value} is done through controlled flooding, in which data packet is forwarded from one node to the next node. However, *an intermediate node is permitted to accept the packet only if the Level ID of the sender node is greater and Sector ID has a difference of utmost 1*. Hence, the Levelling and Sectoring Protocol prevents the unnecessary transmission [5].

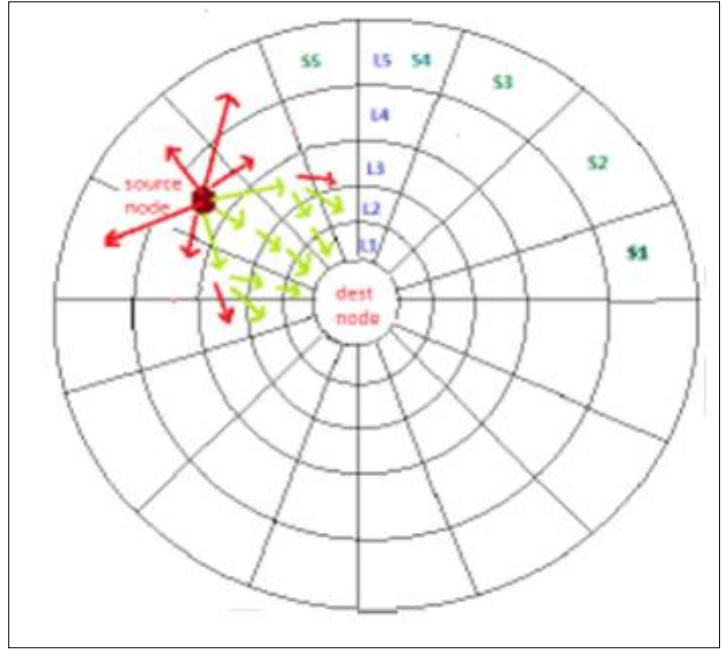

**Fig. 1.** Routing with Levelling & Sectoring (Source: (Chaganty, Murthy, Chilamkurti, & Rho, 2013))

## 3      Motivation

As discussed in the previous section, the supply of energy is restrained with the power supply of the sensor nodes. Therefore, it has become important that the available energy is used in a judicial manner in order to get the desired output from WSNs. Such an efficient use of limited available energy would add up to the cooperative effort and processing capabilities of sensor nodes. The power consumption in a WSN has been attributed to communication and computation operations [7]. As a result of the broadcasting operation in WSNs, unnecessary transmissions take place which lead on to the involuntary involvement of a number of nodes. Energy value associated with the involuntary involved nodes result in wastage of energy in the network. However, with the implementation of Levelling and Sectoring protocol such energy wastage can be minimized. Hence, the motivation of the paper is to estimate the energy savings by evaluating the energy consumption in transmission and reception in the scenarios where no Levelling and Sectoring has been done. The performance of the protocol has been proved in terms of network longevity [5]. The mathematical modelling of energy wastage would help in proving the performance of the protocol in terms of energy efficiency.

## 4    Proposed Work

A WSN constitutes: sensor nodes, processing elements, base station and the interconnection network.

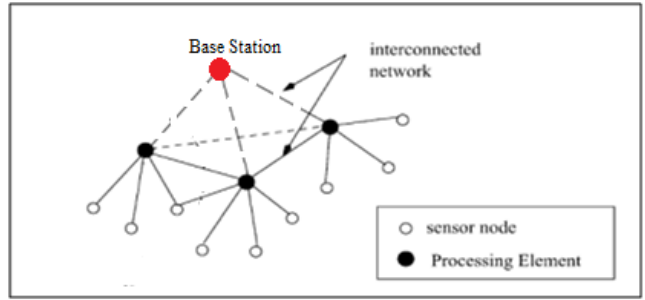

**Fig. 2.** Architecture of Wireless Sensor Network

Network topology plays an important role in the execution of network operations and the related power consumption. In order to realize the wastage, we first consider linear arrangement of nodes (being the simplest arrangement of nodes). In a linear arrangement, there are 'n' sensor nodes placed in a line, with the first node functioning as the Base Station. Assuming the 'k$^{th}$' node to be broadcasting, mathematically

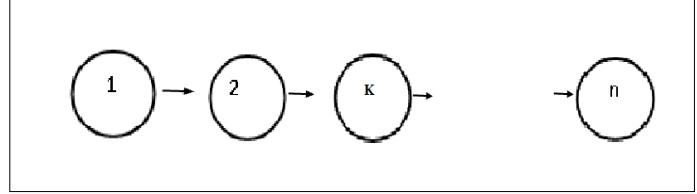

**Fig. 3.** Linear Arrangement of Wireless Sensor Nodes

Number of nodes unnecessarily involved in transmission,

$$B_k = n - k - 1$$

Wasted Transmission Energy,

$$T_x = E_t + E_t(n - k - 1)$$

where, $E_t$ – Transmission energy

Number of nodes unnecessarily involved in reception,

$$B_k = n - k$$

Wasted Reception Energy,
$$R_x = E_r(n - k)$$
where, $E_r$ - Reception energy

## 4.1 Analytical Modelling

If we consider an arbitrary graph, a spanning tree can be extracted which is minimally a binary tree. Hence, we consider the case of balanced binary tree, nested tree and Q-ary tree for mathematically estimating the wastage of energy in transmission/reception. The consideration of simplest network topology would help in computation of energy wastage in absence of Levelling and Sectoring Protocol. The same can be applied to the real time WSNs.

- The WSN is in the form of a Balanced Binary tree of depth 'd'. The root node of the tree is the Base Station (BST). A node at depth 'i' broadcasts information which is being automatically received by the 'i+1'$^{th}$ depth and is transmitted further, since leveling and sectoring protocol is not used. As the nodes are arranged in the form of a binary tree $2^i$ nodes are involved in broadcasting. However, the transmission/reception to 'i+1'$^{th}$ depth cannot be stopped. The nodes placed at further depths (depth greater than 'i+1') are involuntarily involved in transmission/reception (the transmission/reception to such nodes can be stopped through levelling and sectoring).

Therefore,
- For, Pure Flooding :
  - Transmission
    - Number of nodes unnecessarily involved in transmission,

$$B_{t_i} = \left\{\sum_{j=i+2}^{d} 2^{j-i-1}\right\} 2^i \quad (1)$$

    - Wasted Transmission Energy ,

$$T_x = E_t + E_t[B_{t_i}]$$

  - Reception
    - Number of nodes unnecessarily involved in reception ,

$$B_{r_i} = \left\{\sum_{j=i+2}^{d} 2^{j-i}\right\} 2^i \quad (2)$$

    - Wasted Reception Energy,

$$R_x = E_r[B_{r_i}]$$

- For, Controlled Flooding
  In the case of controlled flooding, each node which receives a packet, broadcasts the packet with probability 'p' independent of other nodes.
  - Transmission
    - Expected number of nodes unnecessarily involved in transmission,

$$B_{t_i} = \left\{\sum_{j=i+2}^{d} (2p)^{j-i-1}\right\} 2^i$$

- Expected wasted Transmission Energy,

$$T_x = E_t + E_t[B_{t_i}]$$

- Reception
  - Expected number of nodes unnecessarily involved in reception,

  $$B_{r_i} = \{\sum_{j=i+2}^{d}(2p)^{j-i}\}2^i$$

  - Expected wasted Reception Energy,

  $$R_x = E_r [B_{r_i}]$$

- The wireless sensor network is in the form of a Nested Tree. The tree is binary tree till depth 's'. Further, the tree is tertiary tree from 's +1' till 'd' (entire depth of tree).

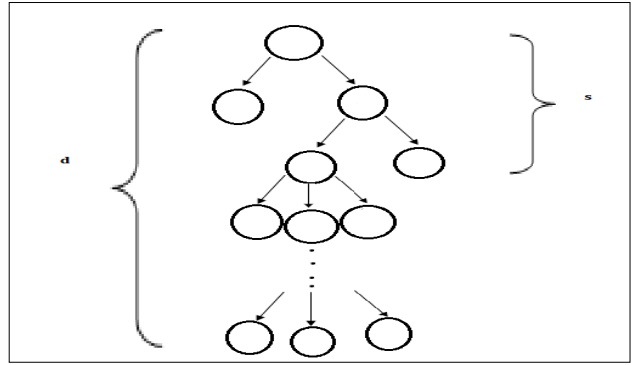

**Fig. 4.** Nested Tree

In the case of nested tree there are two considerations for the value of depth 'i':
- $i > s$
- $i \leq s$

  ➢ Pure Flooding
  ▪ Transmission
    - Case 1: $i > s$
      ♦ Number of nodes unnecessarily involved in transmission,

      $$B_{t_i} = \{\sum_{j=i+2}^{d} 3^{j-i-1}\}3^i \qquad (3)$$

      ♦ Wasted Transmission Energy,

      $$T_x = E_t + E_t[B_{t_i}]$$

- Case 2 : $i \leq s$
  - Number of nodes unnecessarily involved in transmission,
  $$B'_{t_i} = 2^i \left(\sum_{j=i+2}^{s} 2^{j-i}\right) + 2^s \left(\sum_{k=1}^{d-s} 3^{k-1}\right) \quad (4)$$
  - Wasted Transmission Energy,
  $$T_x = E_t + E_t[B'_{t_i}]$$

- Reception
  - Case 1: $i > s$
    - Number of nodes unnecessarily involved in reception,
    $$B_{r_i} = \left\{\sum_{j=i+2}^{d} 3^{j-i}\right\} 3^i \quad (5)$$
    - Wasted Reception Energy,
    $$R_x = E_r[B_{r_i}]$$
  - Case 2 : $i \leq s$
    - Number of nodes unnecessarily involved in reception,
    $$B'_{r_i} = 2^i \left(\sum_{j=i+2}^{s} 2^{j-i}\right) + 2^s \left(\sum_{k=1}^{d-s} 3^{k}\right) \quad (6)$$
    - Wasted Reception Energy,
    $$R_x = E_r[B'_{r_i}]$$

➤ Controlled Flooding
In the case of controlled flooding, each node which receives a packet, broadcasts the packet with probability 'p' independent of other nodes.

- Transmission
  - Case 1: $i > s$
    - Expected number of nodes unnecessarily involved in transmission,
    $$B_{t_i} = \left\{\sum_{j=i+2}^{d} (3p)^{j-i-1}\right\} 3^i$$
    - Expected wasted Transmission Energy,
    $$T_x = E_t + E_t[B_{t_i}]$$
  - Case 2 : $i \leq s$
    - Expected number of nodes unnecessarily involved in transmission,
    $$B'_{t_i} = 2^i \left(\sum_{j=i+2}^{s} (2p)^{j-i}\right) + 2^s \left(\sum_{k=1}^{d-s} (3p)^{k-1}\right)$$

- ◆ Expected wasted Transmission Energy,

$$T_x = E_t + E_t[B'_{t_i}]$$

- Reception
  - Case 1: $i > s$
    - ◆ Expected number of nodes unnecessarily involved in reception,

$$B_{r_i} = \left\{\sum_{j=i+2}^{d} (3p)^{j-i-1}\right\} 3^i$$

    - ◆ Expected wasted Reception Energy,

$$R_x = E_r[B_{r_i}]$$

  - Case 2: $i \leq s$
    - ◆ Expected number of nodes unnecessarily involved in reception,

$$B'_{r_i} = 2^i \left(\sum_{j=i+2}^{s}(2p)^{j-i}\right) + 2^s \left(\sum_{k=1}^{d-s}(3p)^k\right)$$

    - ◆ Expected wasted Reception Energy,

$$R_x = E_r[B'_{r_i}]$$

- The wireless sensor network is in the form of a Q-ary tree of depth 'd'. The following expressions are generalizations of those derived for binary tree.
  - ➢ Pure Flooding
    - Transmission
      - Number of nodes unnecessarily involved in transmission,

$$B_{t_i} = \left\{\sum_{j=i+2}^{d} q^{j-i-1}\right\} q^i$$

      - Wasted Transmission Energy,

$$T_x = E_t + E_t[B_{t_i}]$$

    - Reception
      - Number of nodes unnecessarily involved in reception,

$$B_{r_i} = \left\{\sum_{j=i+2}^{d} q^{j-i}\right\} q^i$$

      - Wasted Reception Energy,

$$R_x = E_r[B_{r_i}]$$

> Controlled Flooding
> In the case of controlled flooding, each node which receives a packet, broadcasts the packet with probability 'p' independent of other nodes.
>  - Transmission
>    - Expected number of nodes unnecessarily involved in transmission,
>
> $$B_{t_i} = \left\{\sum_{j=i+2}^{d}(qp)^{j-i-1}\right\}q^i$$
>
>    - Expected wasted Transmission Energy,
>
> $$T_x = E_t + E_t[B_{t_i}]$$
>
>  - Reception
>    - Expected number of nodes unnecessarily involved in reception,
>
> $$B_{r_i} = \left\{\sum_{j=i+2}^{d}(qp)^{j-i}\right\}q^i$$
>
>    - Expected wasted Reception Energy,
>
> $$R_x = E_r[B_{r_i}]$$

## 5 RESULTS

This section presents the simulation results (performed using MATLAB) for binary tree and nested tree. In order to perform simulations (based on the mathematical modeling), certain assumptions are made for different variable values:-

— As $E_t \gg E_r$, we assumed Transmission Energy, $E_t = 100\ mJ$
— Reception Energy, $E_r = 5\ mJ$

Further, calculations and results have been presented for different depth values of tree.

- Binary Tree
  Let, depth 'i', which is broadcasting information = 2.

The depth values have been considered for three cases (based on difference/interval of depth values considered):

- Case 1 : Interval =1
- Case 2: Interval = 2
- Case 3: Interval =5

The following figure shows the graphs obtained through simulation. The graphs depict the increase in Total Number of Nodes (involuntarily involved in transmission and reception) and Total Energy Wasted (in mJ) with the increase in depth of binary tree for different cases. The graphs also present the constant output for the scenario in which Levelling and Sectoring protocol has been implemented. In this case, the information goes only to 'i+1'$^{th}$ depth. The other depth values ('i+2'$^{th}$, 'i+3'$^{th}$...d) drop the packet as per the Levelling and Sectoring Protocol.

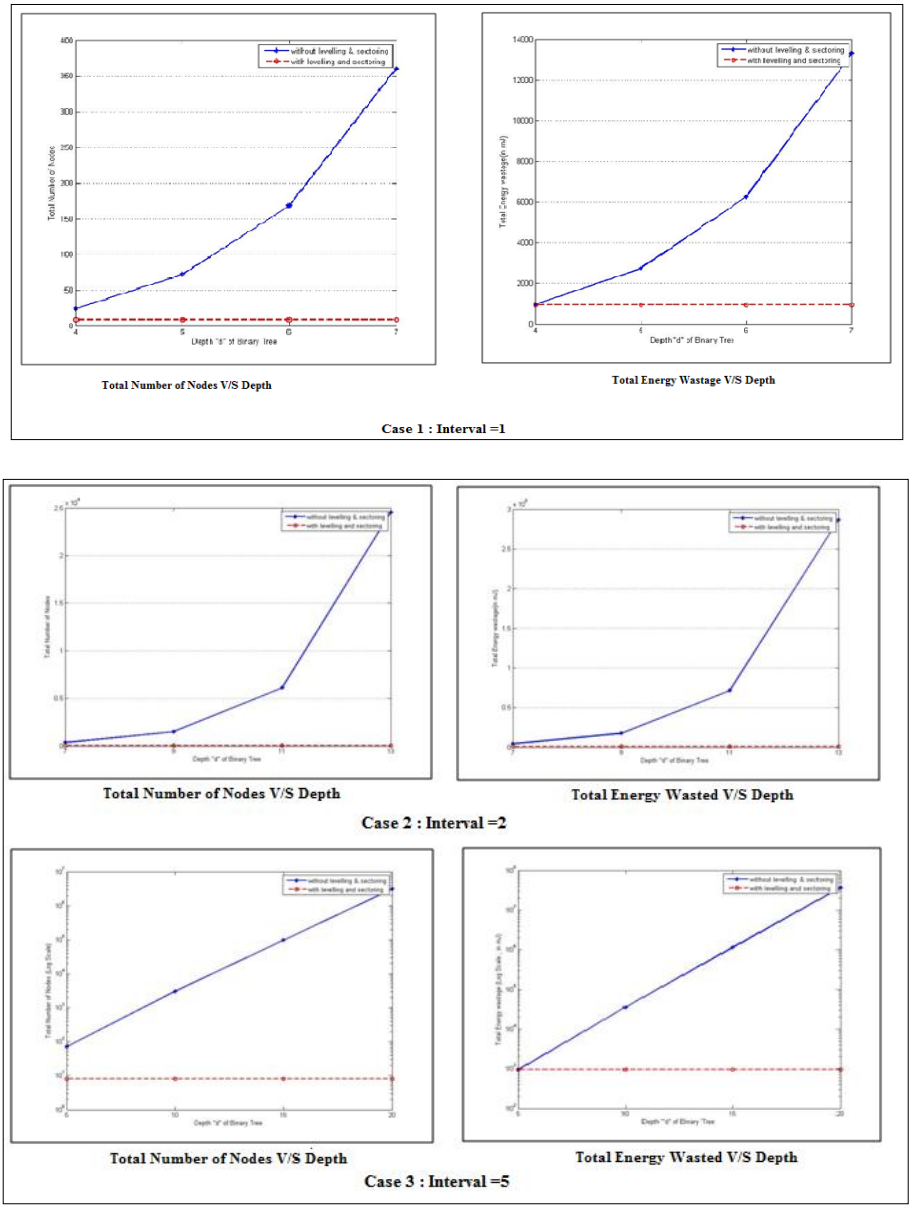

**Fig. 5.** Graphs showing Total Number of Nodes unnecessarily involved & Total Energy Wastage for different depth values.

The following table shows values of total number of nodes and total energy wastage in binary tree for different cases.

Table 1. Results for Different Cases of Binary Tree

| Depth 'd' | Total Number of Nodes $N_i = B_{t_i} + B_{r_i}$ | Total Energy Wasted (in mJ) $E = T_x + R_x$ |
|---|---|---|
| Case 1 : | | |
| 4 | 24 | 980 |
| 5 | 72 | 2740 |
| 6 | 168 | 6260 |
| 7 | 360 | 13300 |
| Case 2 : | | |
| 7 | 360 | 13300 |
| 9 | 1512 | 55540 |
| 11 | 6120 | 224500 |
| 13 | 24552 | 900340 |
| Case 3 : | | |
| 5 | 72 | 2740 |
| 10 | 3048 | 111860 |
| 15 | 98280 | 3603700 |
| 20 | 3145704 | 115342580 |

### 5.1 Nested Tree

The results have been calculated for the two cases as discussed in the mathematical modelling

- $i > s$

Let, the value of depth 'i' which is broadcasting = 3
Depth of binary tree, 's' = 2
The following figure shows the graphs obtained through simulation. The depth values have been considered for three cases (based on difference/interval of depth values considered):

- Case 1 : Interval =1
- Case 2: Interval = 2
- Case 3: Interval = 5

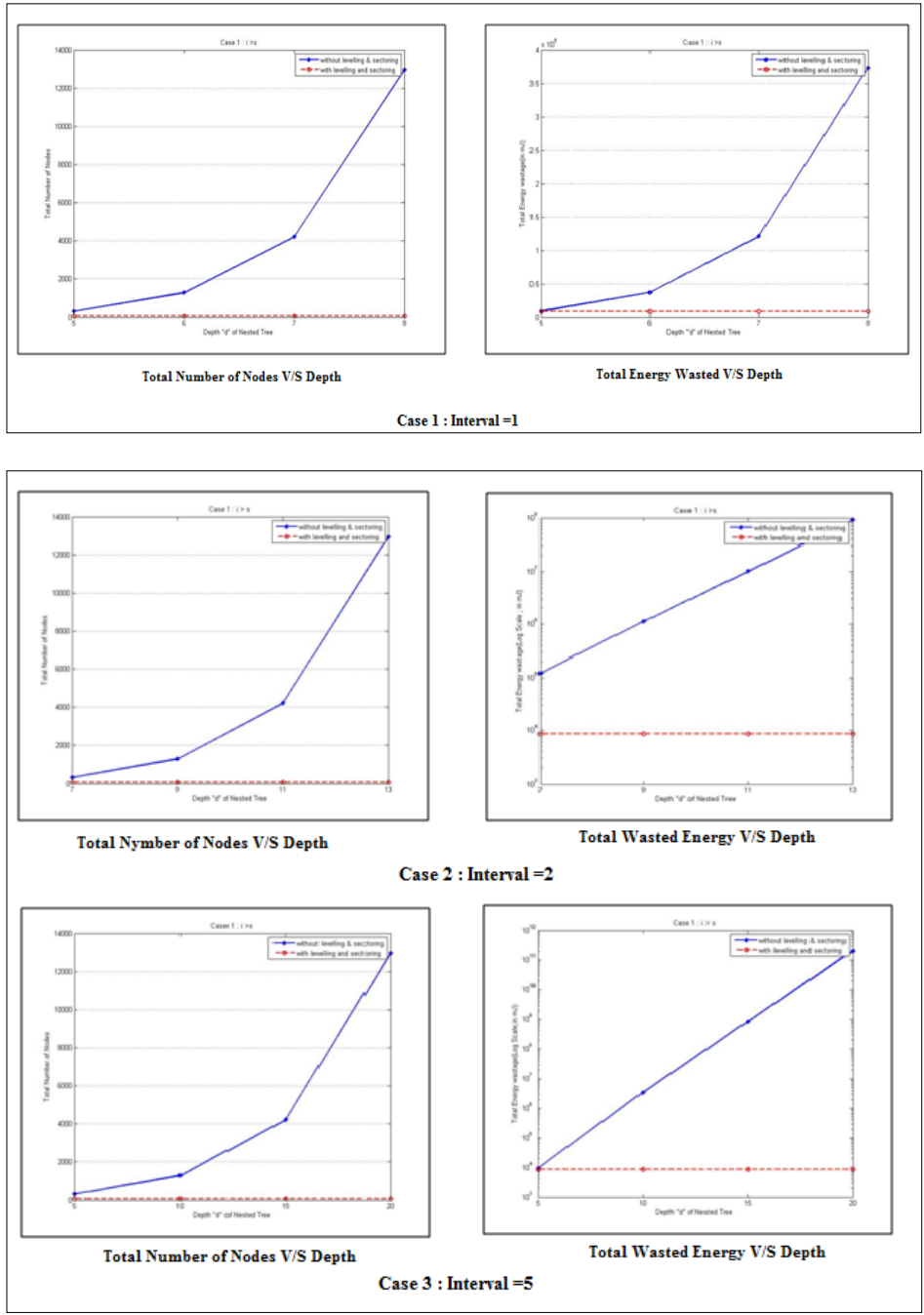

**Fig. 6.** Graphs showing Total Number of Nodes unnecessarily involved & Total Energy Wastage for different depth values

- $i \leq s$

Let, the value of depth 'i' which is broadcasting = 2
Depth of binary tree, 's' = 4

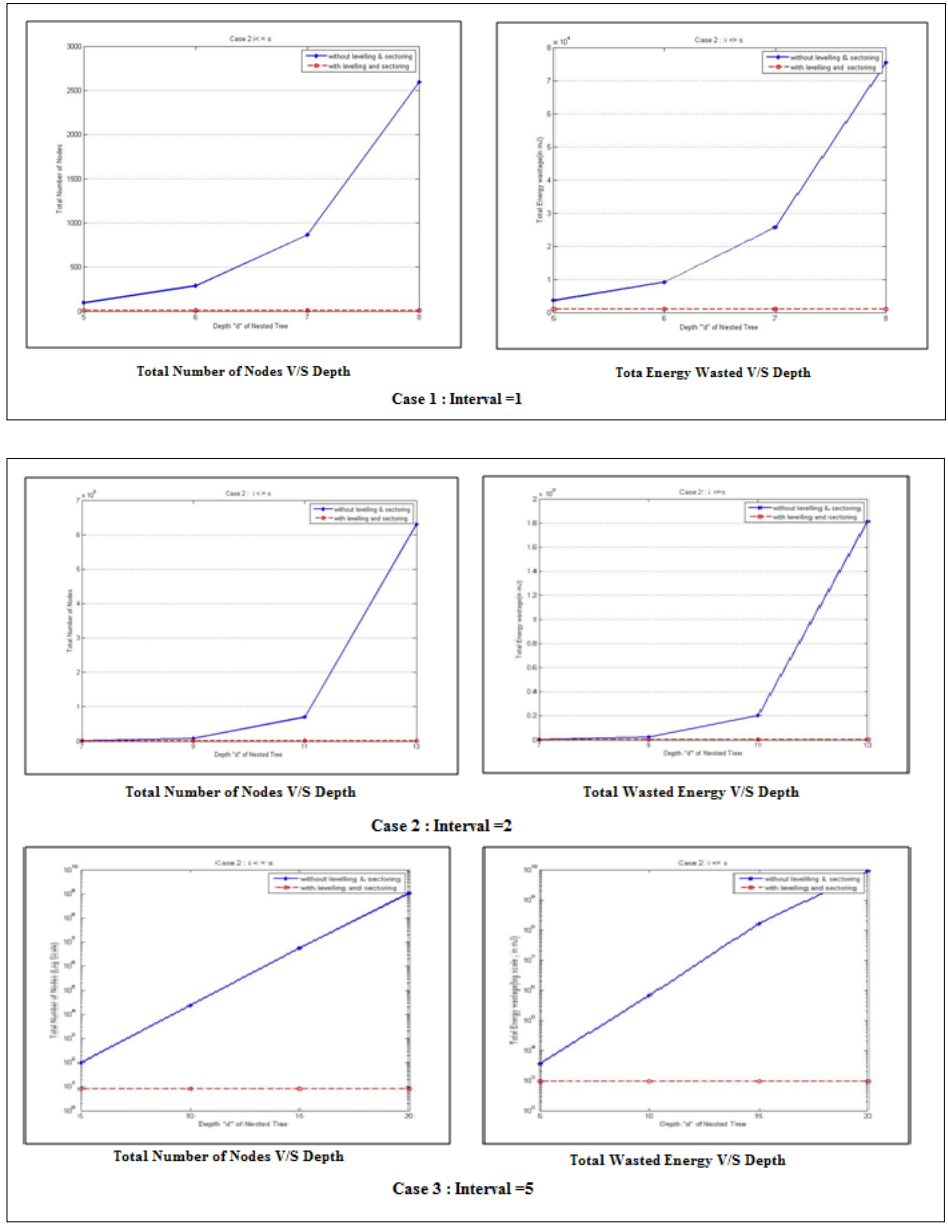

**Fig. 7.** Graphs showing Total Number of Nodes unnecessarily involved & Total Energy Wastage for different depth values

The graphs depict the increase in Total Number of Nodes (involuntarily involved in transmission and reception) and Total Energy Wasted (in mJ) with the increase in depth of nested tree for different cases. The graphs also present the constant output for the scenario in which Levelling and Sectoring protocol has been implemented. In this case, the information goes only to 'i+1'[th] depth. The other depth values ('i+2'[th], 'i+3'[th]...d) drop the packet as per the Levelling and Sectoring Protocol.

The following table shows values of total number of nodes and total energy wastage in nested tree for different cases.

Table 2. Results for Different Cases of Nested Tree

| Depth 'd' | Total Number of Nodes $N_i = B_{t_i} + B_{r_i}$ | | Total Energy Wasted (in mJ) $E = T_x + R_x$ | |
|---|---|---|---|---|
| Case 1 : | $i > s$ | $i \leq s$ | $i > s$ | $i \leq s$ |
| 5 | 324 | 64 | 9415 | 3620 |
| 6 | 1296 | 208 | 37360 | 9140 |
| 7 | 4212 | 640 | 121195 | 25700 |
| 8 | 12960 | 1936 | 372700 | 75380 |
| Case 2 : | | | | |
| 7 | 4212 | 640 | 121195 | 25700 |
| 9 | 39204 | 5824 | 1127215 | 224420 |
| 11 | 354132 | 52480 | 10181395 | 2012900 |
| 13 | 3183484 | 472384 | 91669015 | 18109220 |
| Case 3 : | | | | |
| 5 | 324 | 64 | 9415 | 3620 |
| 10 | 117936 | 17486 | 3390760 | 671540 |
| 15 | 28697652 | 4251520 | 825057595 | 162976100 |
| 20 | 6973568640 | 1033121296 | 200490098500 | 8991982580 |

## 6      Conclusion

The flexibility, low cost, dynamic deployment along with other characteristics has made WSNs realizable for a large number of applications. The consideration of the limitations of the WSNs and a search for solutions for such limitations will develop new aspects of implementation of WSNs. As per the estimations done for different values of depth of a balanced binary tree and nested tree, the results clearly reflect that the number of nodes unnecessarily involved and the total energy which is being consumed increases exponentially with the increase in depth of the tree. The values which have been considered for the depth of the tree are considerably low (in compar-ison to the real time implementation). However, the wastage is significantly high and would further increase with the increase in the number of nodes. Hence, this explains the energy consumption in a wireless sensor network, where levelling and sectoring has not been employed. On the other hand, when Levelling and Sectoring Protocol is taken into consideration the values obtained are low and constant (even with the increase in depth of tree). Hence, the Levelling and Sectoring protocol would help in achieving reliable routing of data, ensuring efficient use of energy. Earlier, network longevity has been proved for the same. In future, the idea of levelling and sectoring can be extended and evaluated for Mobile AdHoc Networks (MANETs).